\documentclass[12pt]{iopart}

\usepackage{iopams}
\usepackage{graphicx}

\newcommand{\re}{\mathrm{Re}}
\newcommand{\im}{\mathrm{Im}}

\begin{document}

%\twocolumn[ %% activate for two-column option

\title{The analogy between optical beam shifts and quantum weak measurements}

\author{Mark R Dennis$^1$ and  J\"org B G\"otte$^{1,2}$}

\address{$^1$H H Wills Physics Laboratory, University of Bristol, Tyndall Avenue, Bristol BS8 1TL, UK\\
$^2$Max-Planck-Institute for the Physics of Complex Systems, Noethnitzer Str. 38, 01187 Dresden, Germany}
\ead{mark.dennis@physics.org}

\begin{abstract}
We describe how the notion of optical beam shifts (including the spatial and angular Goos-H\"anchen shift and Imbert-Federov shift) can be understood as a classical analogue of a quantum measurement of the polarization state of a paraxial beam by its transverse amplitude distribution.
Under this scheme, complex quantum weak values are interpreted as spatial and angular shifts of polarized scalar components of the reflected beam.
This connection leads us to predict an extra spatial shift for beams with a radially-varying phase dependence.
\end{abstract}

\eqnobysec

\section{Introduction}

There are many analogies between phenomena in quantum theory and in classical wave optics, including the connection between Heisenberg's uncertainty principle and the bandwidth theorem \cite{gabor}, transverse optical polarization and spin half as 2-state systems \cite{Aspect}, and the Sch\"odinger equation and paraxial equation of light \cite{kogelnik}, connecting the time propagation of quantum wavepackets and the propagation of narrow, coherent light beams.
Although the underlying physics differs in these cases, the similarity in the underlying mathematics gives rise to analogous phenomena.

Our aim in this paper is to describe and explore the strong analogy between the spatial and angular shifts that light beams experience on reflection from a planar interface, and the notion of weak quantum measurement, incorporating quantum weak values of operators \cite{Aharonov+:PRL60:1988, AharonovRohrlich:WVCH:2005,Mitchinson+}.
Such a connection was identified by Hosten and Kwiat \cite{HostenKwiat:SCI:2008} in the experimental measurement of the spin Hall effect of refracted light beams via quantum weak values.
Here, we generalize this approach, and show that the shift of a paraxial light beam, and its polarized components, is determined by the appropriate average value of an `Artmann operator' related to the reflection matrix for the beam.

Spatial beam shifts are usually small: they are typically comparable to an optical wavelength, which is much smaller than the physical width of the paraxial beam, even when focused on the reflecting interface.
Angular shifts of the direction of the beam, when they exist, are also small, of the order of the (narrow) spectral width of the beam in Fourier space.
The effect originates physically from the Fresnel coefficients' dependence on incidence angle -- as the plane wave components making up the beam deviate from their mean, the complex amplitude and possibly polarization of the reflected plane wave components varies.
The magnitude and direction of the beam shift thus depends on the incidence angle of the beam and its incident polarization, and while we concentrate in this paper on reflection, the case of transmission is very similar.
We will not derive all beam shift formulas here, as they involve some technical details unimportant to the connection to quantum measurements.
For these details (in an explicitly optical approach), we refer the reader to a companion paper \cite{GoetteDennis:NJP:2012}.

Historically, the first beam shift to be discovered was for light beams polarized linearly in, or perpendicular to, the plane of incidence ($p$- and $s$-polarizations respectively), by Goos and H\"anchen in 1943 \cite{GoosHaenchen:AndP436:1947}.
In this case, the shift is in the plane of incidence (referred to as `longitudinal'), given by the famous Artmann formula \cite{Artmann:AndP437:1948}.
If the incident beam has circular polarization (or, more generally, has any polarization not linear in or perpendicular to the plane of incidence), there is an `Imbert-Federov shift' transverse to the plane of incidence, which has been explained in terms of spin-orbit coupling \cite{BliokhBliokh:PRL96:2006}.
%If the polarization of the incident beam is neither fully within the plane of incidence or fully orthogonal to it, there is an `Imbert-Federov' shift transverse to the plane of incidence, which has been explained in terms of spin-orbit coupling \cite{BliokhBliokh:PRL96:2006}.
Goos-H\"anchen and Imbert-Federov shifts are largest in the regime of total reflection, close to the critical angle; when reflection is partial (reflection by a denser medium, or inside the critical angle), Fourier filtering of the spectrum leads to an angular shift of the propagation direction of the light beam \cite{RaBertoniFelsen, AntarBoerner, ChanTamir:OL10:1985}, which has been the subject of much recent experimental attention \cite{Merano+:NatPhot3:2009, Merano+:PRA82:2010}.

\begin{figure}
\begin{center}
\includegraphics[width=9cm]{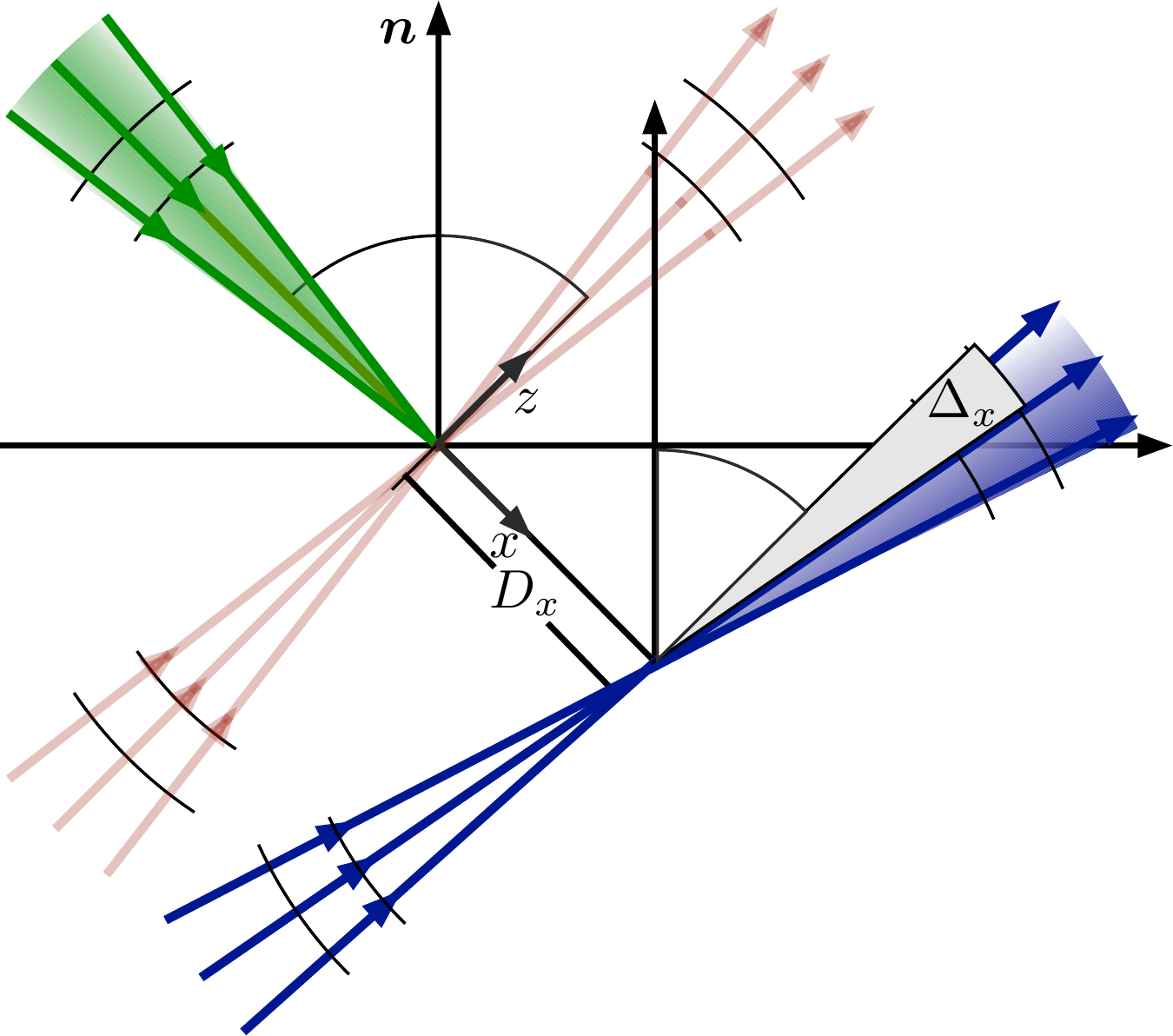}
\caption{\label{fig:shift}
   Schematic illustration of a spatial shift $D_x$ and angular shift $\Delta_x$ of a reflected beam in the plane of incidence.
   If reflection of the incident beam (green) was purely specular, one would observe the `virtual beam' (red).
   However, due to angle-dependence of the reflection coefficients, the real beam is displaced both in position and propagation direction (blue).
 }
\end{center}
\end{figure}

A schematic showing the geometry of a longitudinal shift, both spatial and angular, is shown in Fig.~\ref{fig:shift}.
It is natural to introduce the concept of a `virtual beam', represented in red in the figure.
Inspired by the method of images, this is an ideal, unphysical reflected beam whose position and direction is as if the reflection were purely specular.
Rather than refer to spatial and angular shifts with respect to the incident beam, we instead denote the shifts with respect to the expected position and direction of the virtual beam.
It therefore suffices to work in coordinates based on the virtual beam, which propagates in the $z$-direction, and the plane of incidence is the $xz$-plane, as shown in the figure.
This avoids introducing complicated 3D geometry for the reflection with respect to the incident beam.

Of course, the usual Goos-H\"anchen and Imbert-Federov shifts apply to the centroid of the intensity of the light.
Since the reflected beam also acquires a weakly inhomogenous polarization, analyzing the reflected beam with a polarizer reveals further shifts of polarized components, such as the transverse spin Hall effect of light \cite{BliokhBliokh:PRL96:2006, HostenKwiat:SCI:2008} of circular components when linear polarization is incident; in this case, each circular component undergoes a shift even when there is no net transverse shift to the overall beam.

A particular interest of beam shifts is that the nature and magnitude of the resulting shifts are independent of the spatial profile of the light beam provided the incident beam is axisymmetric and focused on the interface. 
When this condition is relaxed, for instance with a beam with an azimuth-dependent phase vortex $\exp(\rmi \ell \phi)$ \cite{Allen+:PRA45:1992, Dennis+:PO53:2009}, there is an additional `vortex-induced shift,' where the spatial shift involves a combination of the Artmann expressions for the spatial and angular shifts \cite{Bliokh+:OL34:2009, Merano+:PRA82:2010}.

Each of the notions above corresponds to a similar notion in quantum measurement via a weak von Neumann interaction, and our purpose in this paper is to make these explicit.
In the optical reflection case, the interaction is between the polarization and spatial structure of the beam, and the weakness follows from paraxiality (i.e.~localization in Fourier space).
Before describing the details of the analogy, we review the relevant notions from quantum measurements, largely following the description and notation of the very clear exposition of the quantum case by Jozsa \cite{Josza:PRA76:2007}.

\section{Weak quantum measurement}\label{sec:weak}

The weak quantum measurement of an observable $A$ of a quantum system in initial state $|\psi_{\mathrm{i}} \rangle$ is achieved by considering the entire system in a product state of the system to be measured, together with the state of the measurement pointer $|\varphi\rangle,$ i.e.~$|\psi_{\mathrm{i}}\rangle|\varphi\rangle.$
The interaction between the measured system and the pointer is weak, in the sense that the interaction hamiltonian $H_{\mathrm{int}} \equiv g A p,$ acting at a single instant in time, where $g \ll 1$ is a small coupling constant, and $p$ is the momentum operator generating translations of the pointer state $|\varphi\rangle.$
The unitary evolution operator corresponding to the interaction, in units with $\hbar = 1,$  is
\begin{equation}
   \exp(-\rmi H_{\mathrm{int}}) \approx 1 - \rmi g A p,
   \label{eq:unitaryapprox}
\end{equation}
following from the fact that the coupling constant is small, i.e.~the interaction is weak.

Now, the measurement may be made in the usual von Neumann sense, where the measured system's freedoms are `traced over', by taking the (sub-)inner product with $|\psi_{\mathrm{i}}\rangle$, giving rise to the expectation value
\begin{equation}
   \langle A \rangle \equiv \frac{\langle \psi_{\mathrm{i}} | A | \psi_{\mathrm{i}}\rangle}{\langle \psi_{\mathrm{i}} | \psi_{\mathrm{i}}\rangle}.
   \label{eq:expa}
\end{equation}
Alternatively, the projection of the measured system in a certain postselected final state $|\psi_{\mathrm{f}}\rangle$ may be considered, giving rise to the so-called `weak value' of the operator \cite{Aharonov+:PRL60:1988, AharonovRohrlich:WVCH:2005}
\begin{equation}
   A_w \equiv \frac{\langle \psi_{\mathrm{f}} | A | \psi_{\mathrm{i}}\rangle}{\langle \psi_{\mathrm{f}} | \psi_{\mathrm{i}}\rangle}.
   \label{eq:wea}
\end{equation}
When $A$ is hermitian, $\langle A \rangle$ is necessarily real, although it is not for more general $A.$
However, for any $A,$ the weak value $A_w$ is usually complex-valued, and depends on both the pre- and postselected states.

Furthermore, the weak value may take on very large values (`superweak' \cite{BerryShukla:JPA45:2012}), particularly when $|\psi_{\mathrm{i}}\rangle$ and $|\psi_{\mathrm{f}}\rangle$ are almost orthogonal (so the denominator of Eq.~(\ref{eq:wea}) becomes vanishingly small).
We will denote the average value of $A,$ whether the expectation or weak value, by the generally complex 
\begin{equation}
   a \equiv \frac{\langle \psi | A|\psi_{\mathrm{i}}\rangle}{\langle \psi | \psi_{\mathrm{i}}\rangle},
\end{equation} 
where $\langle \psi |$ is $\langle \psi_{\mathrm{i}} |$ or $\langle \psi_{\mathrm{f}} |.$

The final pointer state, after the weak interaction and possibly postselection, is therefore
\begin{equation}
\fl   \langle \psi | \exp(-i H_{\mathrm{int}}) |\psi_{\mathrm{i}}\rangle|\varphi \rangle 
   \approx  \langle \psi |\psi_{\mathrm{i}} \rangle \left( 1 - \rmi g a p\right)|\varphi\rangle 
   \approx  \langle \psi |\psi_{\mathrm{i}} \rangle \exp(-\rmi g a p)|\varphi\rangle. 
\end{equation}
The result of the measurement is a shift in the mean position of the pointer wavefunction $\langle q | \varphi \rangle = \varphi(q) \equiv |\varphi(q)| \rme^{\rmi \chi(q)},$ in the position representation, where $\chi$ is the wavefunction's phase.
When $a$ is real, this simply corresponds to a translation of $\varphi(q)$ by the small amount $g a;$ when $a$ is complex, the expectation value of $q$ is shifted by 
\begin{equation}
    D = g \mathrm{Re}(a) - g \mathrm{Im}(a) \int \rmd q\, q^2 J'(q) 
    = g \mathrm{Re}(a) + 2 g \mathrm{Im}(a) \langle q \chi'(q) \rangle,
    \label{eq:jtheorem1}
\end{equation}
where the second term is nonzero when the pointer wavefunction $\varphi(q)$ at the moment of interaction has a varying phase (so the probability current $J(q) = |\varphi|^2 \chi'$ is nonzero), and we have assumed the unshifted expectation of position $\langle q \rangle = 0.$
The second line here follows from integration by parts, assuming $\chi$ is sufficiently well behaved; in \cite{Josza:PRA76:2007}, the term proportional to $g \im(b)$ was found to be $m \partial_t \langle q^2 \rangle$ by the continuity equation for probability and Schr\"odinger's equation. 
In Fourier space, the pointer's conjugate wavefunction $\widetilde{\varphi}(p)$ also undergoes a shift to its mean position \cite{Steinberg:PRA52:1995, Mitchinson+}, by 
\begin{equation}
    \Delta = g \langle p^2 \rangle \mathrm{Im} (a),
\end{equation}
proportional to the imaginary part of $a$ and the width (variance) of the Fourier transform $\widetilde{\varphi}(p).$
We note that it is common in the quantum mechanical literature to refer to the weak interaction between measured and pointer systems as a `weak measurement', referring to the small magnitude of the coupling constant $g;$ the measurement is weak regardless of whether the weak or expectation value of the operator is being measured.

\section{The analogy between beam shifts and quantum measurement}\label{sec:analogy}

We are now in a position to draw the analogy between beam shifts and quantum weak measurements.
As previously in classical optics analogies with weak values \cite{Ritchie+:PRL66:1990, HostenKwiat:SCI:2008}, the polarization of the beam, represented by the constant transverse vector $\boldsymbol{E},$ is identified with the measured system, and the complex amplitude of the beam corresponds to the pointer.
The measured system is therefore effectively a 2-state system (i.e.~a 2-dimensional Jones vector), and the pointer wavefunction (a normalized, position-dependent complex amplitude) $\varphi(\boldsymbol{r})$ depends on $\boldsymbol{r} = (x,y)$ in the plane perpendicular transverse to the propagation in $z.$ 
A homogeneously polarized light beam, in the quantum language, therefore corresponds to the product state. 
Free paraxial propagation is analogous to free quantum time evolution according to the Schr\"odinger equation.
The paraxial approximation applied to the beam follows from its strong localization around a mean propagation direction in direction (Fourier) space.
A regular or weak measurement of the beam is made depending on whether the centroid of the total intensity of the final reflected beam, or a polarized component of it, are considered.

When a \emph{plane wave} encounters a planar dielectric interface, it is reflected \cite{Jackson:JohnWiley:1998}: it changes direction according to the law of specular reflection, and its polarization $\boldsymbol{E}$ undergoes reflection according to the appropriate Fresnel coefficients $r_s$ and $r_p$ (evaluated at the incidence angle): the $p$-direction perpendicular to the plane of incidence, and the $s$-direction parallel to it, are distinguished as eigenpolarizations of the incident polarization $\boldsymbol{E}.$
The resulting polarization is therefore $\mathbf{R}\cdot\boldsymbol{E},$ where $\mathbf{R}$ is a reflection matrix, diagonal in the $s,p$ basis with entries given by the appropriate reflection coefficients.
It is therefore natural to choose the beam coordinates to correspond to these eigenpolarization directions: $x$ corresponding to $p,$ in the plane of reflection, and $y$ perpendicular to it ($s$ polarization), as in Fig.~\ref{fig:shift}.

When the homogeneously polarized \emph{beam} -- rather than simply a plane wave -- encounters a planar dielectric interface, it approximately undergoes the same change: it is specularly reflected, and its polarization changes according to its central wavevector component.
However, the small variation of wavevector directions about the propagation direction gives rise to a small variation in the incidence angle and plane of incidence of each Fourier component, and it is this small variation, ignored in the approximation of beam as plane wave, that corresponds to weak quantum interaction.

This is made explicit by the introduction of the virtual beam, that is, the imaginary beam propagating in the $z$-direction (i.e.~after specular reflection) as discussed above.
The virtual beam is simply specularly reflected, and its polarization is constant; the `weak interaction' is then the residual small correction (to first order) from the angle dependence. 
It is convenient to perform the analysis in the coordinates of the virtual beam.

Spherical angles for beam coordinates will be denoted by azimuth $\alpha$ and colatitude $\delta,$ i.e.~the transverse wavevector (whose length $k$ is fixed)
\begin{equation}
   \boldsymbol{K} = (K_x,K_y) = k \sin\delta \, (\cos\alpha,\sin\alpha)  \approx k \delta \, (\cos\alpha,\sin\alpha),
\end{equation}
since we assume the spread of the beam in Fourier space is small: paraxiality implies that the variance $\langle \delta^2 \rangle \equiv \frac{1}{k^2}\int\rmd^2\boldsymbol{K}\, |\boldsymbol{K}|^2 |\widetilde{\varphi}(\boldsymbol{K})|^2 \ll 1.$
The polarization of the virtual beam is assumed to be uniform, and determined by $\overline{\mathbf{R}},$ the reflection matrix of the mean wavevector.
$\overline{\mathbf{R}}$ is the reflection matrix for a plane wave incident at $\theta_0,$ the incidence angle of the centre of the beam.

The virtual beam is therefore in a product state, whose amplitude distribution is the same as the initial beam, and whose homogeneous polarization is given by $\boldsymbol{E}_{\mathrm{i}} \equiv \overline{\mathbf{R}}\cdot\boldsymbol{E}.$
It is this virtual beam which plays the role of the preselected quantum product state: the main change to the beam on reflection is its specular change in direction and plane wave reflection-like change to its polarization: the smaller shifts in mean position and direction follow from the weak $\boldsymbol{K}$-dependent variation in $\mathbf{R}.$

The physically reflected beam, in its Fourier representation, is found by multiplication of the virtual beam Fourier transform with the appropriate $\boldsymbol{K}$-dependent reflection matrix $\mathbf{R}.$
Each $\boldsymbol{K}$ has its own particular plane of incidence, and $\mathbf{R}$ applies the appropriate reflection coefficients in the local $s,p$ basis.
Since the spread of $\boldsymbol{K}$-directions around the propagation direction is small, the beam is sensitive only to a low-order Taylor expansion of this matrix $\mathbf{R}$ with respect to $\delta,$ that is
\begin{equation}
   \mathbf{R} \approx \overline{\mathbf{R}} + \delta \, (\cos\alpha,\sin\alpha)\cdot(\overline{\mathbf{R}_x},\overline{\mathbf{R}_y}),
\end{equation}
where $\overline{\mathbf{R}_x}$ and $\overline{\mathbf{R}_y}$ represent the mean derivatives of $\mathbf{R}$ in the longitudinal and transverse directions respectively, evaluated at the mean wavevector.
In this paper, we will not perform the extra geometrical calculations required to find explicit forms of $\overline{\mathbf{R}_x}$ and $\overline{\mathbf{R}_y};$ we derive these in the companion paper \cite{GoetteDennis:NJP:2012} in detail.

It should be clear by analogy with the exposition in Section \ref{sec:weak} above that in this approximation, the reflection operator may be rewritten directly as the action of the mean $\overline{\mathbf{R}}$ followed by an interaction operator entangling the position and polarization.
This can be represented by the mean reflection $\overline{\mathbf{R}}$ left-multiplied by a weak interaction operator (as described above in Section \ref{sec:weak}),
\begin{equation}
   \mathbf{R} \approx \left(\boldsymbol{1} - \rmi \boldsymbol{K}\cdot(\mathbf{A}_x,\mathbf{A}_y) \,\right)\overline{\mathbf{R}} 
   \approx \exp\left(-\rmi \boldsymbol{K}\cdot(\mathbf{A}_x,\mathbf{A}_y)\,\right) \overline{\mathbf{R}},
\end{equation}
where $(\mathbf{A}_x,\mathbf{A}_y)$ is a vector of $2\times 2$ matrices we call `Artmann operators',
\begin{equation}
   \mathbf{A}_j \equiv \frac{\rmi}{k}\overline{\mathbf{R}_j} \, \overline{\mathbf{R}}^{-1}, \quad j = x, y.
\end{equation}
If the reflection is total (i.e.~the reflection coefficients are unimodular), it is straightforward to see (but omitted here) that each $\mathbf{A}_j$ matrix is hermitian.
However, if the reflection is partial, some of the incident light is lost through transmission, resulting in the $\mathbf{A}_j$ matrices being nonhermitian.
The corresponding evolution operator is nonunitary as it does not preserve normalization, as some light is refracted.

In Fourier space, the reflection operator acts as an impulsive evolution of the virtual beam under the `interaction hamiltonian' $\boldsymbol{K}\cdot(\mathbf{A}_x,\mathbf{A}_y),$ which is weak since the transverse momentum $\boldsymbol{K}$ is necessarily small for the paraxial beam.
Apart from the generalization to two dimensions (which is straightforward unless $\varphi(\boldsymbol{r})$ is complex, for which see Section \ref{sec:complex}), the connection with the quantum measurement described above is immediate: the centre of the physical beam in real space, and its direction in Fourier space, are effectively a weak measurement-like shift to the pointer $\varphi(\boldsymbol{r}),$ and its Fourier transform $\widetilde{\varphi}(\boldsymbol{K})$ by the real and imaginary parts of the (possibly complex-valued) average of the Artmann operators $\mathbf{A}_x, \mathbf{A}_y.$

In the case of optical polarization, `tracing out' the polarization degrees of freedom is simply considering the overall intensity of the beam ignoring the polarization; in this case, the spatial shift in $j = x,y$ is given by the general shift (for a real $\varphi(\boldsymbol{r})$)
\begin{eqnarray}
   D_j & = & \re \frac{\boldsymbol{E}_{\mathrm{i}}^* \cdot \mathbf{A}_j \cdot\boldsymbol{E}_{\mathrm{i}}}{\boldsymbol{E}_{\mathrm{i}} \cdot \boldsymbol{E}_{\mathrm{i}}} 
    =  \re \left( \frac{\rmi}{k}\frac{\boldsymbol{E}^*\cdot \overline{\mathbf{R}}^{\dagger} \overline{\mathbf{R}_j} \overline{\mathbf{R}}^{-1}\overline{\mathbf{R}}\cdot\boldsymbol{E}}{\boldsymbol{E}_{\mathrm{i}} \cdot \boldsymbol{E}_{\mathrm{i}}}\right) \nonumber \\
   & = & -\im \left( \frac{1}{k}\frac{\boldsymbol{E}\cdot \overline{\mathbf{R}}^{\dagger} \overline{\mathbf{R}_j} \cdot\boldsymbol{E}}{\boldsymbol{E}\cdot \overline{\mathbf{R}}^{\dagger}\overline{\mathbf{R}} \cdot \boldsymbol{E}}\right).
   \label{eq:spatial}
\end{eqnarray}
When $\boldsymbol{E}$ is linearly polarized in the $x$- or $y$-directions ($p$ and $s$ polarized respectively), the corresponding longitudinal Goos-H\"anchen shift $D_x$ is given by the famous Artmann formulas $-\frac{1}{k}\im r'_p/r_p, -\frac{1}{k}\im r'_s/r_s$ \cite{Artmann:AndP437:1948}, which follow directly from the spatial shift formula in Eq.~(\ref{eq:spatial}), and $D_y = 0$ in this case.
When $\boldsymbol{E}$ is circularly polarized (right- $(+)$ or left- $(-)$ handed), there is a transverse Imbert-Federov shift $D_y$ given by $\pm \frac{1}{k}|r_s+r_p|^2 \cot\theta_0$ \cite{imbert,fedorov, BliokhBliokh:PRL96:2006, Aiello, GoetteDennis:NJP:2012}.
It can be shown that $\overline{\mathbf{R}_y},$ originating from spin-orbit coupling, is a constant times the Pauli matrix $\boldsymbol{\sigma}_1$ in the $s,p$ basis, so in fact $D_y$ is zero if $\boldsymbol{E}$ is linearly polarized.

If reflection is total so the $\mathbf{A}_j$ are hermitian, these spatial shifts are the only shifts which occur.
However, if reflection is partial, there is a shift to the Fourier transform $\tilde{\varphi}(\boldsymbol{K}),$ which is $k$ times an angular shift $\Delta_j$ to the direction of the beam in the $x$ and $y$-directions.
From above, this is proportional to the variance of the beam's Fourier distribution $\langle |\boldsymbol{K}|^2 \rangle = k^2 \langle \delta^2 \rangle,$ where $\langle \delta^2 \rangle$ is the variation of the beam in direction space.
Thus
\begin{eqnarray}
   \Delta_j = & = & \frac{\langle |\boldsymbol{K}|^2 \rangle}{k} \im \frac{\boldsymbol{E}_{\mathrm{i}}^* \cdot \mathbf{A}_j \cdot\boldsymbol{E}_{\mathrm{i}}}{\boldsymbol{E}_{\mathrm{i}} \cdot \boldsymbol{E}_{\mathrm{i}}} \nonumber \\
   & = & \langle \delta^2 \rangle \re \left( \frac{\boldsymbol{E}\cdot \overline{\mathbf{R}}^{\dagger} \overline{\mathbf{R}_j} \cdot\boldsymbol{E}}{\boldsymbol{E}\cdot \overline{\mathbf{R}}^{\dagger}\overline{\mathbf{R}} \cdot \boldsymbol{E}}\right) 
   \label{eq:angular}
\end{eqnarray}
This equation is derived purely from the optical viewpoint in \cite{GoetteDennis:NJP:2012}.
Despite the recent interest in the angular shift, the simple form here in terms of the variance in direction space has not previously been emphasized.
The universality, coming from the relation with weak interactions (although the value of the nonhermitian Artmann operator is not weak here) is one of the main new statements in optical beam shifts in this paper.

\begin{figure}
\begin{center}
\includegraphics[width=12cm]{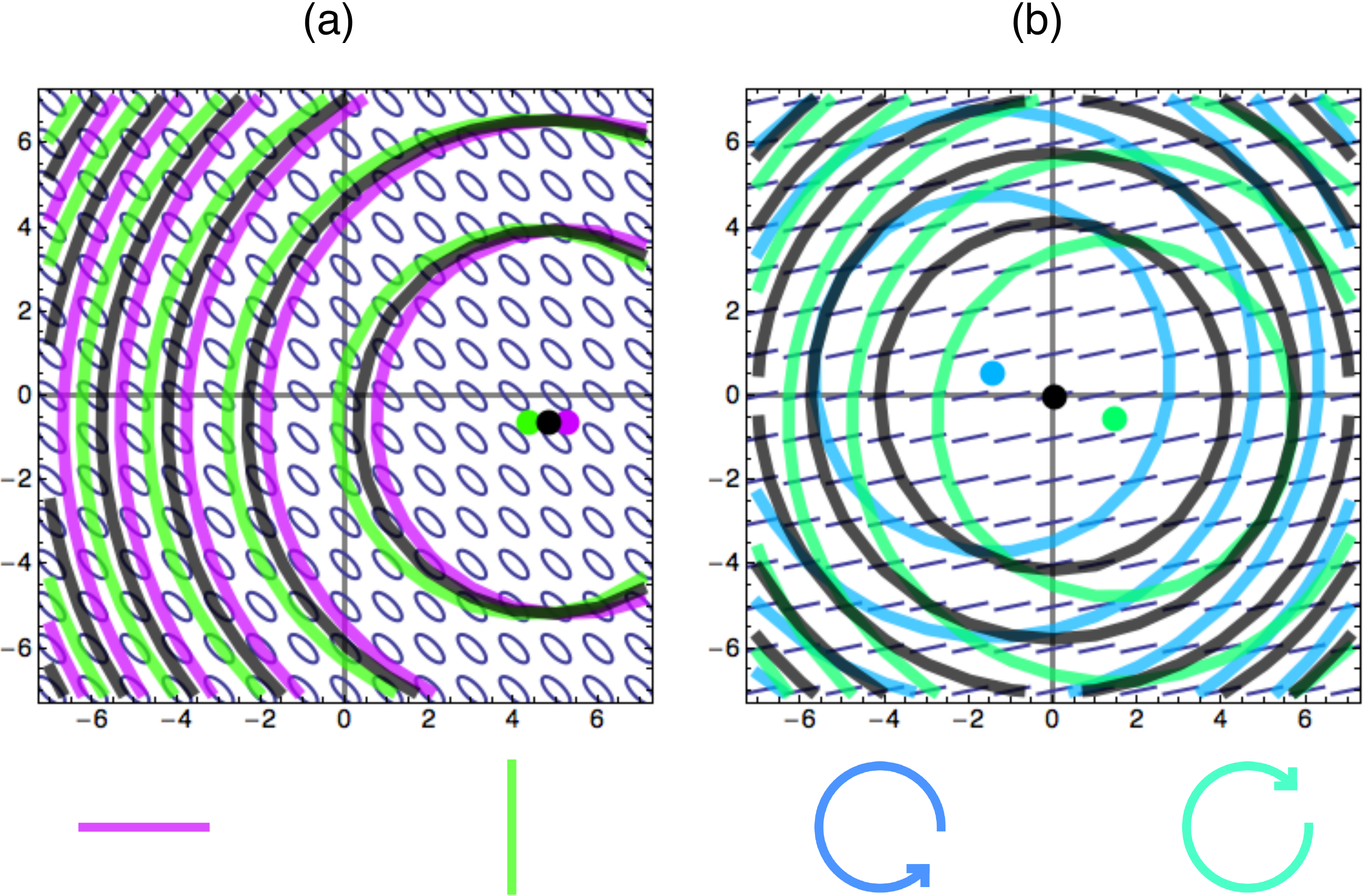}
\caption{\label{fig:polarized}
   Examples of shifted, reflected beams with weakly varying polarization, in units of $k^{-1}.$
   In each case, the incident beam has polarization $\boldsymbol{E} = \frac{1}{\sqrt{2}}(1,1)$, with an incident gaussian profile with width $100k^{-1}$ at incidence angle $\theta_0 = \arcsin 3/4.$ 
   Reflection is between air and glass, with (a) total reflection ($n = 3/2$), (b) partial reflection ($n = 2/3$).
   The plots, in virtual beam coordinates, show contours of overall intensity (grey), and (a) magnitude of components in $p$- and $s$-polarization (purple and green respectively); (b) magnitude of right-hand and left-hand circular components (cyan and turquoise respectively).
   Despite the beam shifts being different, the variation in polarization from the virtual beam polarization $\boldsymbol{E}_0$ is imperceptible at this scale (blue ellipses and lines).
 }
\end{center}
\end{figure}

What of weak values themselves?
As discussed above, postselection is clearly here represented by the presence of a polarizing analyzer $\boldsymbol{F},$ which projects the beam onto a `final' polarization state.
In terms of optical physics, the analogue of the entanglement between measured system and pointer is a position-dependent polarization pattern: the mean position of different final polarization states varies (Fig.~\ref{fig:polarized}), depending on the details of the incident polarization and reflection matrix.
The component shift, or `weak' shift, can therefore be written as follows: the component spatial shift is 
\begin{equation}
   D_{wj} = \re \frac{\boldsymbol{F}^* \cdot \mathbf{A}_j \cdot\boldsymbol{E}_{\mathrm{i}}}{\boldsymbol{F} \cdot \boldsymbol{E}_{\mathrm{i}}} 
   = -\frac{1}{k} \im \left( \frac{\boldsymbol{F}^*\cdot \overline{\mathbf{R}_j} \cdot\boldsymbol{E}}{\boldsymbol{F}^*\cdot \overline{\mathbf{R}} \cdot \boldsymbol{E}}\right),
   \label{eq:wspatial}
\end{equation}
again for $j = x, y,$ and the component angular shift is
\begin{equation}
   \Delta_{wj} = \frac{\langle |\boldsymbol{K}|^2 \rangle}{k} \im \frac{\boldsymbol{F}^* \cdot \mathbf{A}_j \cdot\boldsymbol{E}_{\mathrm{i}}}{\boldsymbol{F}^* \cdot \boldsymbol{E}_{\mathrm{i}}} 
   = \langle \delta^2 \rangle \re \left( \frac{\boldsymbol{F}^*\cdot \overline{\mathbf{R}_j} \cdot\boldsymbol{E}}{\boldsymbol{F}^*\cdot \overline{\mathbf{R}} \cdot \boldsymbol{E}}\right) .
   \label{eq:wang}
\end{equation}
Of course, the weak shift typically has an angular contribution even when the $\mathbf{A}_j$ are hermitian, representing the inhomogeneity of the polarization pattern in direction space, and weak shifts can be rather different directions to their strong counterparts (or exist when these are zero, as is the case of the spin Hall effect for light \cite{HostenKwiat:SCI:2008}).

An illustration of optical beam shifts, including component shifts, is shown in Fig.~\ref{fig:polarized}.
In each case, on the lengthscale of $k^{-1},$ the inhomogeneous polarization is extremely weak, although both longitudinal and transverse shifts are present.
A systematic theoretical study of the component and beam shifts for different choices of $\boldsymbol{E}$ and $\boldsymbol{F}$ is made in \cite{GoetteDennis:NJP:2012}.

For certain choices of $\boldsymbol{F},$ the component shift may be very large, with a rather small overall amplitude.
However, it is important to note that such superweak shifts occur when $\boldsymbol{F}^*\cdot \boldsymbol{E}_{\mathrm{i}} = \boldsymbol{F}^*\cdot \overline{\mathbf{R}} \cdot \boldsymbol{E}$ is small, and the inner product between initial and final polarizations $\boldsymbol{F}^*\cdot \boldsymbol{E}$ does not directly play a role.
This is of course because the analogue of the prepared quantum state is the virtual beam.

These optical shifts completely mirror quantum mechanical pointer shifts, and the 2-dimensional nature does not change any of the essential description.
However, if the complex pointer amplitude $\varphi(\boldsymbol{r})$ has radial- and azimuthal-varying complex phase, the situation is more complicated than the one-dimensional case described in \cite{Josza:PRA76:2007}, and we describe this in the following section.

\section{The case of radial and azimuthally varying pointer phase}\label{sec:complex}

In this section, we will represent the weak or expectation value of the Artmann operator by the complex 2-dimensional vector $\boldsymbol{a} = (a_x,a_y),$ which is defined by one of the following:
\begin{equation}
   a_j = \frac{\boldsymbol{E}\cdot \overline{\mathbf{R}}^{\dagger} \overline{\mathbf{R}_j} \cdot\boldsymbol{E}}{\boldsymbol{E}\cdot \overline{\mathbf{R}}^{\dagger}\overline{\mathbf{R}} \cdot \boldsymbol{E}}
   \quad\hbox{or}\quad \frac{\boldsymbol{F}^*\cdot \overline{\mathbf{R}_j} \cdot\boldsymbol{E}}{\boldsymbol{F}^*\cdot \overline{\mathbf{R}} \cdot \boldsymbol{E}}, \quad j = x,y.
\end{equation}
Although our main example is optical beams, the discussion applies to any weakly interacting quantum system with two noncommuting operators $A_x, A_y$ multiplying the two components of pointer momentum, and our aim is to generalize the contribution of $\im(a)$ to the spatial pointer shift.

Following \cite{Josza:PRA76:2007}, we consider the modulus squared (intensity or probability distribution) of the reflected wave after tracing out or postselecting the polarization state,
\begin{equation}
   |(1 - \rmi \boldsymbol{K}\cdot \boldsymbol{a})\varphi(\boldsymbol{r})|^2 \approx |\rme^{-\re(\boldsymbol{a})\cdot\nabla}\varphi|^2 - \rmi \, \im(\boldsymbol{a})\cdot (\varphi^* \nabla \varphi - \varphi \nabla \varphi^*)
   \label{eq:jshift}
\end{equation}
where $\boldsymbol{K},$ as the momentum, corresponds to the generator of translations $-\rmi \nabla,$ and only terms up to the first order in $\boldsymbol{a}$ are kept as usual.
The first term represents the intensity pattern shifted by $\re(\boldsymbol{a})$, and the second term represents additional interference caused between the differently-weighted Fourier components with varying incident phases.

This second term generalizes the second term on the RHS of (\ref{eq:jtheorem1}) to 2-dimensional pointer wavefunctions; the first term here -- a pure translation -- is just the usual spatial shift.
The shift is the mean of $\boldsymbol{r}$ with respect to this distribution, and we assume that the mean position of the unshifted (virtual) beam is the origin.

We begin our analysis of the second term of Eq.~(\ref{eq:jshift}) by rewriting it as the (optical or probability) current vector $\boldsymbol{J} \equiv -\frac{\rmi}{2}(\varphi^*\nabla\varphi - \varphi\nabla\varphi^*),$
\begin{equation}
   - \rmi \, \im(\boldsymbol{a})\cdot (\varphi^* \nabla \varphi - \varphi \nabla \varphi^*) = 2 \,\im(\boldsymbol{a})\cdot \boldsymbol{J}.
\end{equation}
Its contribution to the $x$-component of the spatial shift is therefore
\begin{equation}
   D_x^{\mathrm{extra}} \equiv 2 \int \rmd^2 \boldsymbol{r}\, x \,\im(\boldsymbol{a})\cdot \boldsymbol{J} = -\int \rmd^2 \boldsymbol{r}\, x^2 \partial_x\left[\im(\boldsymbol{a})\cdot \boldsymbol{J}\right],
   \label{eq:shiftcorr}
\end{equation}
with the equality following from Green's theorem.
An equivalent expression holds for the shift in the $y$-direction.

In the one-dimensional case \cite{Josza:PRA76:2007}, this is related through conservation of current and the Schr\"odinger equation to the rate of change of the variance of the wavefunction as in Eq.~(\ref{eq:jtheorem1}).
This is not possible in two or higher dimensions since the integral does not involve the divergence of $\boldsymbol{J},$ and the general situation appears rather more complicated.

However, in most optical fields of interest, any varying phase factor factorizes into an azimuthal part (say with a vortex with quantum number $\ell$) and a radial part, i.e. for azimuthal angle $\phi$ and radius $r,$
\begin{equation}
   \varphi(\boldsymbol{r}) = \sqrt{I(r)} \exp(\rmi \ell \phi + \rmi f(r) ),
   \label{eq:phaseansatz}
\end{equation}
for intensity $I(r) = |\varphi|^2,$ and $f(r)$ some radius-dependent phase function.
It should be noted that, for optical reflection, the azimuthal index $\ell$ reverses sign on reflection \cite{Fedoseyev:OC:2001,leachetal}.
Our convention here is that Eq.~(\ref{eq:phaseansatz}) refers to the virtual beam, so the azimuthal dependence of the incident beam is $\exp(-\rmi \ell \phi).$

It is easy to see that, for the virtual beam (\ref{eq:phaseansatz}) on returning to cartesian cordinates,
\begin{equation}
   \boldsymbol{J} = I(r) \left[ \frac{\ell}{r^2} (-y,x) + \frac{f'(r)}{r} (x,y) \, \right].
\end{equation}
Using this form in the formula for the extra shift in Eq.~(\ref{eq:shiftcorr}) in $x,$ and after integrating out the azimuth $\phi,$ we have
\begin{eqnarray}
\fl   D_x^{\mathrm{extra}} 
   = -\frac{1}{8} \im(\boldsymbol{a})\cdot\int_0^{\infty} \rmd r\, r \left(r I(r)[f'(r)+3r f''(r)]+3r^2 f'(r)I'(r), \ell[3r I'(r)-2I(r)]\right) \nonumber \\
   = \im(\boldsymbol{a})\cdot(\langle r f'(r)\rangle, \ell), 
\end{eqnarray}
where the second line follows from normalization and integration by parts, assuming $f(r)$ is reasonably well-behaved.
An identical argument for $y$ gives
\begin{equation}
   D_y^{\mathrm{extra}} = \im(\boldsymbol{a})\cdot(-\ell,\langle r f'(r) \rangle),
\end{equation}
so the net total shift in two dimensions, analogous to (\ref{eq:jtheorem1}) for a pointer with the assumed form, is
\begin{equation}
   \boldsymbol{D} = \re(\boldsymbol{a}) + (\langle r f'(r) \rangle +\rmi \ell \boldsymbol{\sigma}_2)\cdot\im(\boldsymbol{a}),
\end{equation}
where $\sigma_2$ is the second Pauli matrix.

This is the form of the the spatial shift for complex $\boldsymbol{a}$ and complex $\varphi(\boldsymbol{r}).$
The $\ell$-dependent parts correspond to the `vortex-induced shift' \cite{Fedoseyev:OC:2001, dasguptagupta, Bliokh+:OL34:2009, Merano+:PRA82:2010}: the $\ell$-fold twisting of the phase fronts results in $\im(a_y),$ usually determining the angular shift in $y,$ contributing to the real shift in $x$ weighted by $\ell$ and similarly (in a manner preserving the sense of circulation), $-\ell\im(a_x)$ contributes to the shift in $y.$
Apart from the azimuthal quantum number $\ell,$ no other features of the incident beam contribute to this part of the shift.

The other part, given by $\im(\boldsymbol{a})\langle r f'(r)\rangle,$ directly generalizes the corresponding term in (\ref{eq:jtheorem1}).
In optical beams, a radially-varying phase is interpreted as curved phasefront (rather than a beam at its focus, where the phasefronts are flat and $f(r)=0$), due, for instance, to the Gouy phase \cite{Gouy:CRAS110:1890,Siegman:USB:1986}.  
This radial phase variation leads to a term proportional to the expectation value of phasefront gradient $\langle rf'(r)\rangle$ times the usual angular shift.

A simple example to consider is a gaussian light beam (gaussian wavepacket) of width $w$ propagating according to the paraxial equation, $\varphi(\boldsymbol{r},z) = \exp(-r^2/[w^2(1+ \rmi z/z_{\mathrm{R}})])/(1+\rmi z/z_{\mathrm{R}}),$ where $z_{\mathrm{R}} = k w^2/2,$ the Rayleigh distance.
In this case, $\langle r f'(r) \rangle = z/z_{\mathrm{R}},$ implying the radial part of the extra shift is small when the beam is focused close to the interface; this result naturally is similar to the one-dimensional \cite{Josza:PRA76:2007} constant related to the rate of expansion of the transverse waveform.  
To our knowledge, this additional `defocusing-induced shift' has not been experimentally measured, but is present as an additional feature of a more complicated amplitude pattern, corresponding to a more complicated quantum mechanical pointer wavefunction \cite{AharonovBotero:PRA72:2005, BerryShukla:JPA45:2012}. In particular it is different to a focal shift, that is a shift of the focus along the propagation axis of the beam as discussed in \cite{McGuirkCarniglia:JOSA67:1977}.

\section{Discussion}\label{sec:disc}

We have shown how the reflection of a narrow optical beam is a classical wave analogue to a quantum weak measurement.
Although the experiment could be done for an ensemble of single photons, nothing in our analysis relies strictly on the quantum nature of the light: the Hilbert space is completely classical, describing polarization and (transverse) position of the optical beam, with entanglement of the degrees of freedom interpreted as a position-dependent polarization pattern \cite{SimonGori2010}.

As such, this phenomenon adds to many well-known classical optical analogues of quantum phenomena involving polarization and complex optical amplitudes.
Although our paper highlights the analogy between the weak values and optical beam shifts of polarization components, the discussion applies to the whole beam without a polarizing analyzer, which is itself analogous to a usual expectation value.
In particular, the effect of the weak measurement on the imaginary part of an operator's average value is immaterial of the kind of average (weak or expectation); the shift to the momentum wavefunction and possibly extra spatial shift always occurs for average values of nonhermitian operators, realised here in partial reflection.

The main effect of reflection, of course, is to change a beam's propagation direction and polarization, giving what we call the virtual beam. 
It is the virtual beam which is analogous to the quantum prepared system, whose position and polarization weakly interact, causing it to be shifted in both position and direction.
Rather than coming from a small coupling constant, the weakness comes from paraxiality, that is narrowness of the beam about its mean propagation direction in Fourier space.
In this case, as we have seen, the main change to a beam on reflection is the small shift in position and possibly direction; for nonparaxial beams (such as the dipole radiation field considered in \cite{Berry:PRLSLA467:2011}), the shifts are present, but the reflected field is much more complicated.

The effects described here are only to first order, as are most phenomena studied in the physics of weak measurement.
Second order effects become important for superweak values, whenever the analyzer is almost orthogonal to the polarization of the virtual beam. 
The magnitude of the spatial and angular shifts corresponding to superweak values is explored in an optical setting elsewhere \cite{GoetteDennis:OL37:2012}.
With the exception of the extra shift coming from the pointer wavefunction's varying phase, all of the shifts are independent of the spatial distribution of the amplitude of the beam (assuming its modulus is radially symmetric).
This simplifies previous descriptions of the angular shift in optics, as the shift has a universal form when measured in units of the angular spread $\langle \delta^2 \rangle.$
Higher-order effects involve changes to the shape of the beam, and these will reveal more subtle structures associated with the entanglement between position and polarization.

It should be stressed that the shift of a beam on reflection does not require polarization, only an incidence angle-dependent reflection coefficient $r(\theta_0).$
This might be achieved with acoustic waves at a lossy wall (similar to the partial reflection case), or Robin boundary conditions (similar to total reflection) \cite{DennisGoette:SPIE7950:2011}. 
From the discussion above, it is clear that even in this case there is a complex scalar shift $a = (\rmi/k) (r'/r) ,$ whose real and imaginary parts contribute in the ways described above: the real part gives rise to a longitudinal spatial shift, the imaginary part to a longitudinal angular shift, and possibly to an additional longitudinal spatial shift (if the wave has a radius-dependent phase) and a transverse shift (if there is an azimuthal phase).
Most of the transverse shifts described above are absent as they require spin (polarization) interacting with azimuthal (orbital) terms.

The fact that weak values generalise so simply to classical wave physics cements their significance as physical quantities.
Furthermore, the generality of the beam shift framework as we have outlined suggests generalization to other kinds of waves, not only acoustic waves as suggested above, but also elastic waves and shifts to matter waves \cite{deHaan+:PRL104:2010,balasz}.
This suggests that further analysis of such simple physical phenomena as the reflection of a beam at a dielectric might reveal new insights into quantum physics.

%%?? acknowledgements

\section*{Acknowledgements}

We are grateful to Michael Berry and Kostya Bliokh for valuable comments.
This work was carried out when JG was a Royal Society Newton Fellow.
MRD is a Royal Society University Research Fellow.

%%%
%%% REFERENCES
%%%

\section*{References}

%\bibliographystyle{unsrt}
%\bibliography{weakvalues}

\end{document}